\documentclass[12pt]{article}
\usepackage[latin1]{inputenc}

\usepackage{amsmath}
\usepackage{color}
\usepackage{amsfonts}
\usepackage{amssymb}
\usepackage{graphicx}
\usepackage{geometry}
\usepackage{amssymb,epsfig,subfigure}
\usepackage{hyperref}
\usepackage{comment}
\usepackage{tabularx}
\usepackage{bm}
\usepackage{euscript}
\usepackage{graphicx}
\usepackage{color}
\usepackage{amsfonts}
\usepackage{exscale}
\usepackage{amsbsy}
\usepackage{subfigure}
\usepackage{textcomp}
\usepackage{comment}
\usepackage{hyperref}
\usepackage{slashed}
\usepackage{authblk}

\usepackage{tabularx}
\usepackage{bm}
\usepackage{euscript}
\usepackage{graphicx}
\usepackage{color}
\usepackage{amsfonts}
\usepackage{exscale}
\usepackage{amsbsy}
\usepackage{subfigure}
\usepackage{textcomp}
\usepackage{comment}
\usepackage{hyperref}

\def\be{\begin{equation}}
\def\ee{\end{equation}}
\def\bea{\begin{eqnarray}}
\def\eea{\end{eqnarray}}
\def\ba#1\ea{\begin{align}#1\end{align}}
\def\bg#1\eg{\begin{gather}#1\end{gather}}
\def\bm#1\em{\begin{multline}#1\end{multline}}
\def\bmd#1\emd{\begin{multlined}#1\end{multlined}}

\usepackage[numbers,sort&compress]{natbib}

\def\simge{
    \mathrel{\rlap{\raise 0.511ex 
        \hbox{$>$}}{\lower 0.511ex \hbox{$\sim$}}}}

\def\simle{
    \mathrel{\rlap{\raise 0.511ex 
        \hbox{$<$}}{\lower 0.511ex \hbox{$\sim$}}}}

\makeatletter
\renewcommand\section{\@startsection {section}{1}{\z@}%
                                 {-3.5ex \@plus -1ex \@minus -.2ex}
                                   {2.3ex \@plus.2ex}%
                                   {\normalfont\large\bfseries}}
\renewcommand\subsection{\@startsection{subsection}{2}{\z@}%
                                   {-3.25ex\@plus -1ex \@minus -.2ex}%
                                     {1.5ex \@plus .2ex}%
                                     {\normalfont\bfseries}}
\renewcommand\subsubsection{\@startsection{subsubsection}{3}{\z@}%
                                   {-3.25ex\@plus -1ex \@minus -.2ex}%
                                     {1.5ex \@plus .2ex}%
                                     {\normalfont\itshape}}
\makeatother

\def\pplogo{\vbox{\kern-\headheight\kern -29pt
\halign{##&##\hfil\cr&{\ppnumber}\cr\rule{0pt}{2.5ex}&\ppdate\cr}}}
\makeatletter
\def\ps@firstpage{\ps@empty \def\@oddhead{\hss\pplogo}%
  \let\@evenhead\@oddhead 
}
\def\maketitle{\par
 \begingroup
 \def\thefootnote{\fnsymbol{footnote}}
 \def\@makefnmark{\hbox{$^{\@thefnmark}$\hss}}
 \if@twocolumn
 \twocolumn[\@maketitle]
 \else \newpage
 \global\@topnum\z@ \@maketitle \fi\thispagestyle{firstpage}\@thanks
 \endgroup
 \setcounter{footnote}{0}
 \let\maketitle\relax
 \let\@maketitle\relax
 \gdef\@thanks{}\gdef\@author{}\gdef\@title{}\let\thanks\relax}
\makeatother

\numberwithin{equation}{section}

\begin{document}

\setcounter{page}0
\def\ppnumber{\vbox{\baselineskip14pt
}}

\def\ppdate{
} \date{\today}

\title{\bf  Weiss oscillations and particle-hole symmetry at the half-filled Landau level
\vskip 0.5cm}
\author[1]{Alfred K. C. Cheung}
\author[1,2]{S. Raghu}
\author[3]{Michael Mulligan}
\affil[1]{\small \it Department of Physics, Stanford University, Stanford, CA 94305, USA}
\affil[2]{\small \it SLAC National Accelerator Laboratory, 2575 Sand Hill Road, Menlo Park, CA 94025, USA}
\affil[3]{\small \it Department of Physics and Astronomy, University of California,
Riverside, CA 92511, USA}

\bigskip

\maketitle

\begin{abstract}
Particle-hole symmetry in the lowest Landau level of the two-dimensional electron gas requires the electrical Hall conductivity to equal $\pm e^2/2h$ at half-filling. We study the consequences of weakly broken particle-hole symmetry for magnetoresistance oscillations about half-filling in the presence of an applied periodic one-dimensional electrostatic potential using the Dirac composite fermion theory proposed by Son. At fixed electron density, the oscillation minima are asymmetrically biased towards higher magnetic fields, while at fixed magnetic field, the oscillations occur symmetrically as the electron density is varied about half-filling. We find an approximate ``sum rule" obeyed for all pairs of oscillation minima that can be tested in experiment.
The locations of the magnetoresistance oscillation minima for the composite fermion theory of Halperin, Lee, and Read (HLR) and its particle-hole conjugate agree exactly.
Within the current experimental resolution, the locations of the oscillation minima produced by the Dirac composite fermion coincide with those of HLR.
These results may indicate that all three composite fermion theories describe the same long wavelength physics. 
\end{abstract}
\bigskip

\newpage

\tableofcontents

\vskip 1cm

\section{Introduction}

The low-temperature response to an applied magnetic field can provide important information about the zero-field ground state of a system \cite{Ashcroft}.
For instance, the free electron gas exhibits oscillatory behavior as a function of the inverse applied magnetic field $(1/B)$ \cite{Shoenberg}.
These quantum oscillations, which are found in both thermodynamic and transport quantities, occur symmetrically about $B=0$ when the square of the inverse Fermi wave vector $1/k^2_F$ is commensurate with the square of the magnetic length $\ell_B^2 = c\hbar/e |B|$.
Thus, as a probe of the zero-field ground state, quantum oscillations reveal the most basic quantity of the electron gas, its Fermi wave vector.

Weiss oscillations \cite{Weissfirst, gerhardtsweissklitzing, WinklerKotthausPloog1989} are quantum oscillations that occur because of the presence of a periodic scalar or vector potential. 
The length scale provided by the period of the imposed potential allows additional oscillations to occur when the cyclotron radius is (approximately) commensurate with the period \cite{Weiss1990}.
These oscillations occur at magnetic field values $B(p)$ satisfying
\begin{align}
\label{weissformula}
\ell^2_{B(p)} = {d \over 2 k_F} \Big(p - \phi\Big), \quad p =  1,  2,  3, \ldots, 
\end{align}
where $d$ is the period of the potential, $\phi = 1/4$ ($-1/4$) for a scalar (vector) potential modulation, and we have taken the potential to be periodic along one spatial direction \cite{zhanggerhardts, peetersvasilopoulos1992scalar, peetersvasilopoulosmagnetic, gerhardts1996}.
The most easily resolved oscillation minima are those at small $p$.\footnote{We assume throughout that the applied periodic potential is static. In experimental systems, the periodic potential can sometimes vary over a characteristic timescale $\tau_{\rm pot}$. Consequently, only those minima satisfying the static approximation,  $k_F \ell^2_{B(p)}/v_F \ll \tau_{\rm pot}$, are observed, where $v_F$ is the Fermi velocity.}
When $k_F > 2/d$, the Weiss oscillation minima at small $p$ occur nearer to $B=0$ than the de Haas - van Alphen or Shubnikov - de Haas oscillation minimum that is produced when $\ell_B^2 = 1/k_F^2$.
This is particularly important if the zero-field state is unstable at sufficiently large magnetic fields.  

In a similar fashion, quantum oscillations occur \cite{kangstormerpfeifferbaldwinwest1993, smet1998, smet1999, willett1999, kamburov2012} {\it approximately} symmetrically about half-filling of the lowest Landau level of the two-dimensional electron gas (2DEG) when the electron density $n_e = B/2\Phi_0$ where the quantum of magnetic flux $\Phi_0 = hc/e$.
These observations are surprising from the perspective of the underlying Fermi gas at $B=0$. 
They suggest the possibility of an emergent Fermi liquid-like ground state at half-filling \cite{Jiang1989transportanomalies}.
Indeed, near half-filling, the system admits an alternative description in terms of ``composite fermions" \cite{halperinleeread, kalmeyerzhang, jainCF}, which can heuristically be thought of as bound states of an electron and two vortices.   In the traditional approach of \cite{halperinleeread, kalmeyerzhang, jainCF}, composite fermions ``feel" on average, zero effective magnetic field $b = 2 \Phi_0 n_e - B$ at half-filling and their density remains the same as the electron density $n_e$.
The quantum oscillations found approximately symmetrically about $b = 0$ have been interpreted as a striking confirmation of this phenomenological picture.

Recent experiments \cite{Kamburov2014, LiuDengWaltz} measuring the locations about half-filling of the magnetoresistance minima in the presence of a one-dimensional periodic scalar potential motivate the reexamination of the above picture of composite fermions. In these experiments, the electron density was held fixed while the uniform transverse magnetic field was tuned about its half-filling value. 
It was reported that Weiss oscillation minima above $B^+(p)$ and below $B^-(p)$ half-filling are {\it not} symmetric about $B= 2 \Phi_0 n_e$.
While the inferred density of the excitations contributing to the Weiss oscillations appears to equal the electron density below half-filling, consistent with \cite{halperinleeread}, the density is equal to the hole density $n_h = B/\Phi_0 - n_e$ above half-filling.
Within the composite fermion framework, this result suggests that the nature of the composite fermions changes across half-filling: the composite fermions are ``electron-like" below and ``hole-like" above half-filling \cite{BMF2015}.

Son has proposed a manifestly particle-hole symmetric effective description of the state at half-filling \cite{Son2015}.
(Within the random phase approximation, an electron-like or hole-like composite fermion theory does not exhibit particle-hole symmetry \cite{kivelson1997, BMF2015}, apparently, in contradiction to experiment \cite{Shahar1995, Wong1996}.)
His theory has been argued to obtain from a sort of percolation transition between composite fermion theories that are electron-like (hole-like) below (above) half-filling \cite{mulliganraghufisher2016}.
In Son's theory (reviewed in Sec. \ref{sonstheory}), the composite fermion is a Dirac fermion at a density fixed by the applied magnetic field, rather than the electron or hole density.
This dependence of the Dirac composite fermion density on the applied magnetic field has interesting implications for the predicted Weiss oscillations that we study in this note.

In order to summarize a particular consequence of our results, we highlight the following distinction: it is possible to depart from half-filling by either varying the applied magnetic field with fixed electron density or by varying the electron density at fixed field.
Within Son's theory, the locations of the Weiss minima have a different character depending upon these two possibilities.
In the former case, we find the magnetic ``sum rule":  
\begin{align}
\label{magsumrule}
B^+(p) + B^-(p) = {4 \eta \over (p - \phi)^2 d^2} + 4 \Phi_0 n_e.
\end{align}
where $\eta = c\hbar/e$.
Note that Dirac composite fermions couple to an applied periodic scalar (vector) potential as a periodic vector (scalar) potential.
Consequently, $\phi = -1/4$ ($1/4$) should be substituted into \eqref{magsumrule} when a periodic scalar (vector) potential is applied to the electronic system. This magnetic ``sum rule" is {\it qualitatively} consistent with the observations in \cite{Kamburov2014}. In the latter case, the oscillations occur symmetrically about half-filling as the electron density is varied. Thus, these ``sum rules" provide an interesting test of Son's theory or any description of the half-filled Landau level.

Because Son's theory involves a Dirac fermion, our analysis benefits from prior theoretical work \cite{matulispeetersweiss2007, tahirsabeehmagneticweiss2008, 2016arXiv161004068B} studying Weiss oscillations in graphene. We summarize this work and describe how it can be applied to the Dirac composite fermion theory in Sec. \ref{weissoscillations}. We then discuss the implications of these results for Weiss oscillations near half-filling in Secs. \ref{weaklybroken}. In Sec. \ref{comparison}, we compare the Weiss oscillations produced by the Dirac composite fermion theory with those expected of the composite fermion theories in \cite{halperinleeread, BMF2015}. 
In the regime of parameter space probed by experiment, we find the locations of the magnetoresistance oscillation minima produced by the various composite fermion theories agree.
We summarize and conclude in Sec. \ref{discussion}.

\section{Weiss oscillations of Dirac composite fermions}

\subsection{Dirac composite fermions}
\label{sonstheory}

We now review the Dirac composite fermion theory of the half-filled Landau level.

Electrons in the lowest Landau level near half-filling can be described by an electrically charged (two-component) Dirac fermion $\Psi_e$. 
The benefit of the Dirac formulation is that the limit of vanishing Landau level mixing $\omega_c = eB/mc \rightarrow \infty$ can be smoothly achieved at fixed magnetic field by taking the Dirac electron mass $m\rightarrow 0$.
The resulting Dirac electron lagrangian becomes
\begin{align}
\label{electronshalf}
{\cal L}_{\rm e} = \bar{\Psi}_e \gamma^\mu (i \partial_\mu + A_\mu) \Psi_e + {1 \over 8 \pi} \epsilon^{\mu \nu \rho} A_\mu \partial_\nu A_\rho,
\end{align}
where $A_\mu$ with $\mu \in \{t,x,y\}$ represents the background electromagnetic gauge field, $\bar{\Psi}_e = \Psi_e^\dagger \gamma^t$ with $\gamma$-matrices $\gamma^t = \sigma^3, \gamma^x = i \sigma^1, \gamma^y = i \sigma^2$ where $\sigma^j$ are the Pauli-$\sigma$ matrices, and $\epsilon^{t x y} = 1$.
To simplify the expressions of this section, we set $e = \hbar = c = 1$ and so $\Phi_0 = 2 \pi$; we will restore these constants later when appropriate. In terms of the Dirac electrons, the electron density, \begin{align}
n_e \equiv \frac{ \delta \cal L}{\delta A_t}= \Psi_e^\dagger \Psi_e + {B \over 4 \pi},
\end{align}
where the background magnetic field $B = \partial_x A_y - \partial_y A_x > 0$. Thus, when $n_e = B/4\pi$, i.e., when $\nu \equiv 2\pi n_e/B = 1/2$, the Dirac electrons half-fill the zeroth Landau level. At vanishing Dirac electron mass, i.e., when there is no Landau level mixing, the electron lagrangian is invariant under the anti-unitary ($i \mapsto - i$) particle-hole transformation (with respect to the lowest Landau level) which takes $t \mapsto - t$,
\begin{align}
\Psi_e & \mapsto - \gamma^t \Psi_e^\ast, \cr
(A_t, A_x, A_y) & \mapsto (- A_t, A_x, A_y),
\end{align}
and shifts the lagrangian by a filled Landau level, ${\cal L}_e \mapsto {\cal L}_e + {1 \over 4 \pi} \epsilon^{\mu \nu \rho} A_\mu \partial_\nu A_\rho$. Unbroken particle-hole symmetry at half-filling ensures the zero-temperature limit of the dc electrical Hall conductivity $\sigma_{xy} = {1 \over 2}$ in units of $e^2/h$.

Son conjectured \cite{Son2015} a dual Dirac composite fermion lagrangian whose precise formulation has been refined in \cite{seibergsenthilwangwitten2016} (following a more general duality conjecture \cite{WangSenthilfirst2015, metlitskiviswanathp-v2016}) to
\begin{align}
\label{refineddual}
{\cal L} = \bar{\psi} \gamma^\mu (i \partial_\mu + a_\mu) \psi - {1 \over 8 \pi} \epsilon^{\mu \nu \rho} a_\mu \partial_\nu a_\rho - {2 \over 4 \pi} \epsilon^{\mu \nu \rho} c_\mu \partial_\nu c_\rho + {1 \over 2 \pi} \epsilon^{\mu \nu \rho} c_\mu \partial_\nu (a_\rho + A_\rho).
\end{align}
In \eqref{refineddual}, $\psi$ is the Dirac composite fermion, while $a_\mu$ and $c_\mu$ are dynamical 2+1D gauge fields. For our work, we can simplify \eqref{refineddual} by integrating out $c_\mu$ to find
\begin{align}
\label{simpledual}
{\cal L} = \bar{\psi} \gamma^\mu (i \partial_\mu + a_\mu) \psi + {1 \over 4 \pi} \epsilon^{\mu \nu \rho} a_\mu \partial_\nu A_\rho + {1 \over 8 \pi} \epsilon^{\mu \nu \rho} A_\mu \partial_\nu A_\rho.
\end{align}
In \eqref{simpledual}, the particle-hole transformation acts by taking
\begin{align}
\psi & \mapsto \gamma^y \psi, \cr
(a_t, a_x, a_y) & \mapsto (a_t, - a_x, - a_y), \cr
(A_t, A_x, A_y) & \mapsto (- A_t, A_x, A_y), \cr
\end{align}
and by shifting the lagrangian by a filled Landau level. In the dual frame, the electron density,
\begin{align}
\label{densitydictionary}
n_e = {1 \over 4 \pi} \Big(b + B \Big),
\end{align}
where $b = \partial_x a_y - \partial_y a_x$ is the magnetic flux of the emergent gauge field.
The $a_t$ equation of motion fixes 
\begin{align}
\label{compositedensity}
\psi^\dagger \psi = - {B \over 4 \pi}.
\end{align}
Thus, the Dirac composite fermions are placed at a density that is determined by the external magnetic field. This is to be contrasted with the theories of \cite{halperinleeread, BMF2015} in which the composite fermion density is equal to either the electron or hole density. Loosely speaking, \eqref{electronshalf} and \eqref{simpledual} are {\it fermionic} particle-vortex duals where a background magnetic field is traded for finite charge density and electrons (particles) are replaced by vortices of an emergent gauge field. 

Coulomb interactions and couplings to additional background potentials supplement \eqref{electronshalf} and \eqref{simpledual}. The degeneracy of the half-filled zeroth Landau level of a free Dirac fermion implies that the low-energy physics, i.e., the ground state of the system, is extremely sensitive to the nature of the added interaction. The advantage of the dual formulation \eqref{simpledual} is to provide an apparently non-degenerate starting point from which to consider these interactions.

Upon departing from half-filling, \eqref{densitydictionary} implies that a non-zero emergent magnetic field $b = 4 \pi n_e - B \neq 0$ develops in the Dirac composite fermion theory. There are two canonical ways in which this can occur: by fixing the electron density and varying the magnetic field or by fixing the magnetic field and varying the electron density. In the former case where the magnetic field varies, \eqref{compositedensity} implies that the Dirac composite fermion density likewise varies; in the latter case, the composite fermion density does not change. We consider both cases simultaneously by keeping in mind that the Dirac composite fermion density in general depends on the applied magnetic field.

\subsection{Weiss oscillations of Dirac composite fermions}
\label{weissoscillations}

Next, we study the effects of a periodic scalar or vector potential on the magnetoresistivity of the state described by \eqref{simpledual}.
Thus, we must first understand how the electrical resistivity is related to observables in the Dirac composite fermion theory and how to couple the Dirac composite fermions to a periodic scalar or vector potential. We then turn to a summary for how the periodic potential produces an oscillatory correction to the Dirac composite fermion conductivity.

\subsubsection{Electrical and Dirac composite fermion conductivity}

Linear response relates the measured electrical current $J$ to the applied electric field $E$,
\begin{align}
\label{electricalresponsedef}
J_i = \sigma_{ij} E^j.
\end{align}
To directly relate the electrical conductivity $\sigma_{ij}$ to observables in the Dirac composite fermion theory, we use the relations,
\begin{align}
\label{cDiracurrent}
j_i^\psi & = - {1 \over 4 \pi} \epsilon_{ij} E^j, \\
\label{electricalcurrenttranslation}
J_i & = {1 \over 4 \pi} \epsilon_{ij} (e^j + E^j),
\end{align} 
where $j_i^\psi = \bar{\psi} \gamma^i \psi$, $e^i = \partial^i a^t - \partial^t a^i$, $\epsilon_{xy} = - \epsilon_{yx} = 1$, and $i, j \in \{x,y\}$. The first relation follows from the equations of motion of the spatial components $a_i$ of the dynamical gauge field.  
The second equation follows from the definition of the response current to an externally applied gauge field: $J_i = \delta {\cal L}/\delta A_i$. The Dirac composite fermions couple to the emergent gauge field as electrons couple to electromagnetism. Therefore, linear response with respect to $e_i$ implies
\begin{align}
\label{linresponsecomposite}
j_i^\psi = {1 \over 2 \pi} \sigma_{ij}^\psi e^j,
\end{align}
and defines the dimensionless Dirac composite fermion conductivity $\sigma_{ij}^\psi$. Equating \eqref{cDiracurrent} and \eqref{linresponsecomposite} in order to solve for $e_i$ and then substituting the result into \eqref{electricalcurrenttranslation}, we may read off the linear electrical conductivity from its definition in \eqref{electricalresponsedef}:
\begin{align}
\label{electricalconductivity}
\sigma_{ij} = {1 \over 2} {e^2 \over h} \Big(\epsilon_{ij} - {1 \over 2} \epsilon_{ik} (\sigma^\psi)^{-1}_{kl} \epsilon_{lj} \Big).
\end{align}
When the Dirac composite fermions exhibit vanishing Hall conductivity, the electrical Hall conductivity takes its particle-hole symmetric value $\sigma_{xy} = {1 \over 2} e^2/h$. Furthermore, \eqref{electricalconductivity} says $\sigma_{xx} = {1\over 4} {e^2 \over h} \rho_{yy}^{\psi}$ and $\sigma_{yy} = {1 \over 4} {e^2 \over h} \rho_{xx}^{\psi}$ where $\rho^{\rm \psi}$ is the dimensionless Dirac composite fermion resistivity.

\subsubsection{Coupling to a periodic scalar potential}

A periodic electrostatic scalar potential obtains from a non-zero scalar component of the electromagnetic gauge field,
\begin{align}
\label{scalarpotential}
A_t = V \cos(K x),
\end{align}
where $K = 2\pi/d$ and we assume the bandwidth $V \ll E_F$, the Fermi energy.
The $a_y$ equation of motion obtained from \eqref{simpledual} fixes the $y$-component of the Dirac composite fermion current,
\begin{align}
\bar{\psi} \gamma^y \psi = - {K V \over 4 \pi} \sin(K x).
\end{align}
We accommodate this constraint by turning on a non-zero background component of the emergent vector potential,
\begin{align}
\label{periodicvector}
\vec{a} = \Big(0, \tilde{V} \sin(K x)\Big).
\end{align}
where $\tilde{V}$ vanishes when $V = 0$.
(This is analogous to how a magnetic field generates a non-zero Dirac composite fermion density which may be accommodated by turning on a chemical potential for the composite fermions.) Thus, we see that a periodic electromagnetic scalar potential couples to the Dirac composite fermions as a periodic vector potential.

A periodic magnetic potential can also be considered. By the same reasoning as above, it can be seen that the Dirac composite fermions couple to this magnetic perturbation via a periodic emergent scalar potential. We will not explicitly review this case in the analysis below; instead, we will only quote the result which can be readily found by similar methods that we present next.

\subsubsection{Weiss oscillations of a Dirac fermion}
\label{weissofdirac}

In our approach to the study of Weiss oscillations, we make a crucial simplification to the Dirac composite fermion dynamics: we ignore the fluctuations of the emergent gauge field $a$. Consequently, we only examine the qualitative effects of the Dirac composite fermion picture. 
In Sec. \ref{discussion}, we remark upon the interpretation of this approximation and argue that strong Coulomb interactions (partially) justify it.

After ignoring the fluctuations of the emergent gauge field, our problem simplifies to the calculation of the correction to the conductivity $\sigma_{ij}^\psi$ of a Dirac fermion at density $B$ in the presence of a magnetic field $b = 4 \pi n_e - B$ due to a periodic vector potential \eqref{periodicvector}. Happily, this problem has been considered in \cite{tahirsabeehmagneticweiss2008}; see \cite{matulispeetersweiss2007} for the case of a periodic scalar potential. We now summarize their results and apply them to the case of the Dirac composite fermion.

The Kubo formula for the diffusive component of the (dimensionless) dc Dirac composite fermion conductivity says \cite{charbonneauvVV1982}
\begin{align}
\label{kubo}
\sigma^\psi_{ij} = {1 \over L_x L_y} \sum_{M} \Big(- \partial_{E_M} f_D(E_M)\Big) \tau(E_M) v^M_i v^M_j,
\end{align}
where $L_x$ ($L_y$) is the length of the system in the $x$-direction ($y$-direction), $\beta^{-1} = k_B T$ is the temperature, $M$ denotes the quantum numbers of the single-particle states, $f_D(E) = \Big(1 + \exp(\beta (E - E_F)\Big)^{-1}$ is the Fermi-Dirac distribution function with Fermi energy $E_F > 0$, $\tau(E_M)$ is the scattering time for states at energy $E_M$, and $v_{i,j}^M$ is the velocity in the $i,j$-direction of the state $M$. We stress that our interpretation and use of this formula assumes there are no (relevant) quantum fluctuations of $a$. (We ignore any corrections to the collisional component of the conductivity \cite{charbonneauvVV1982} which are typically suppressed \cite{peetersvasilopoulos1992scalar, peetersvasilopoulosmagnetic} in comparison to \eqref{kubo}.)

We are interested in the contribution to the conductivity that arises from the presence of the periodic potential along the $x$-direction.
Assuming a constant $\tau(E_M) = \tau$ scattering time for all states, we need only calculate the correction to the energies $E_M$ which determine the velocities $v_i^M = \partial_{k_i} E_M$. After choosing an appropriate gauge, we will show that only $v_y^M$ obtains a new contribution (to leading order in $\tilde{V}$) and so we restrict our attention to $\sigma^\psi_{yy}$ or $\rho_{xx}$ using \eqref{electricalconductivity}. 

The (non-interacting) hamiltonian for the composite Dirac fermion is
\begin{align}
H = v_F \vec{\sigma}\cdot (i \vec{\partial} + \vec{a}),
\end{align}
where 
\begin{align}
\vec{a} = \Big(0, b x + \tilde{V} \sin(K x) \Big).
\end{align}
We study the effect of the periodic potential on the $b \neq 0$ spectrum to first order in $\tilde{V}$.
At $\tilde{V} = 0$, the eigenstates form Landau levels indexed by integers $n = 0, 1, 2, \ldots$,
\begin{align}
\psi_{n, k_y}(x,y) = {e^{i k_y y} \over \sqrt{2 L_y \ell_b}} \begin{pmatrix} - i \Phi_{n-1}\Big({x + x_b \over \ell_b} \Big) \cr i \Phi_{n}\Big({x + x_b \over \ell_b} \Big)
\end{pmatrix},
\end{align}
where $k_y \in {2 \pi \over L_y} \mathbb{Z}$ is the momentum along the $y$-direction ($L_y \rightarrow \infty$), $x_b(k_y) \equiv x_b = k_y \ell_b^2$, and $\Phi_n(z) = {e^{- z^2/2} \over \sqrt{2^n n! \sqrt{\pi}}} H_n(z)$ for the $n$-th Hermite polynomial $H_n(z)$.
Thus, the single-particle states can be labeled by the quantum numbers $M = (n, k_y)$.
The energy of these states,
\begin{align}
E_{n, k_y}({\tilde{V} = 0}) \equiv E_{n, k_y}^{(0)} =  \pm v_F {\sqrt{2 n} \over \ell_b}.
\end{align}
Disorder generally lifts the degeneracy of the flat Landau level bands and provides a non-zero contribution to the velocity $v_i^M$.
We have included this contribution in assuming a constant scattering time $\tau$; instead, we are interested in the leading effects of the periodic potential.

First order perturbation theory says the leading correction to the energy,
\begin{align}
E_{n, k_y}^{(1)} & = \tilde{V} \int_{- {L_x \over 2}}^{{L_x \over 2}} dx \int^{L_y \over 2}_{- {L_y \over 2}} dy\ \psi^\dagger_{n, k_y}(x,y) \sin(K x) \sigma^y \psi_{n, k_y}(x,y) \cr
& = {v_F \tilde{V} \sqrt{2 n} \over K \ell_b} \cos(K x_b) e^{- {z}/2} \Big[L_{n-1}(z) - L_{n}(z) \Big],
\end{align}
where $L_n(z)$ with $z = K^2 \ell_b^2/2$ is the $n$-the Laguerre polynomial and we have ignored terms that vanish in the thermodynamic limit $L_x, L_y \rightarrow \infty$.
(For the case of a periodic scalar potential applied to the Dirac composite fermion system, the relative minus sign between the Laguerre polynomials becomes a plus sign and there is no $\sqrt{n}$ Landau level index coefficient.)
We implicitly assume parameters are such that the first-order energy corrections are much less than the Landau level spacings, $|E_{n, k_y}^{(1)}| \ll E_{n, k_y}^{(0)} - E_{n-1, k_y}^{(0)}$. Thus, the leading contribution to the velocity arising from the periodic potential,
\begin{align}
\Delta v_y^{n, k_y} = \partial_{k_y} E_{n, k_y}^{(1)}.
\end{align}
Substituting $v_i^{n, k_y} = \delta_{iy} \Delta v_y^{n, k_y}$ into the Kubo formula \eqref{kubo}, we find the correction to the Dirac composite fermion dc conductivity,
\begin{align}
\label{generalconductivity}
\Delta \sigma_{yy}^{\psi} & = {\tau \over \hbar L_x L_y} \sum_{n = 0}^{\infty} \int^{{L_x \over 2 \ell^2_b}}_{-{L_x \over 2 \ell^2_b}} {dk_y \over 2 \pi/L_y} \Big( - \partial_{E_{n, k_y}} f_D(E_{n, k_y})\Big) \Big(\Delta v_y^{n, k_y}\Big)^2.
\end{align}
The integral over $k_y$ sums over the $L_x L_y/ \ell_b^2$ states within each Landau level $n$.

\begin{figure}
\includegraphics[width=\columnwidth]{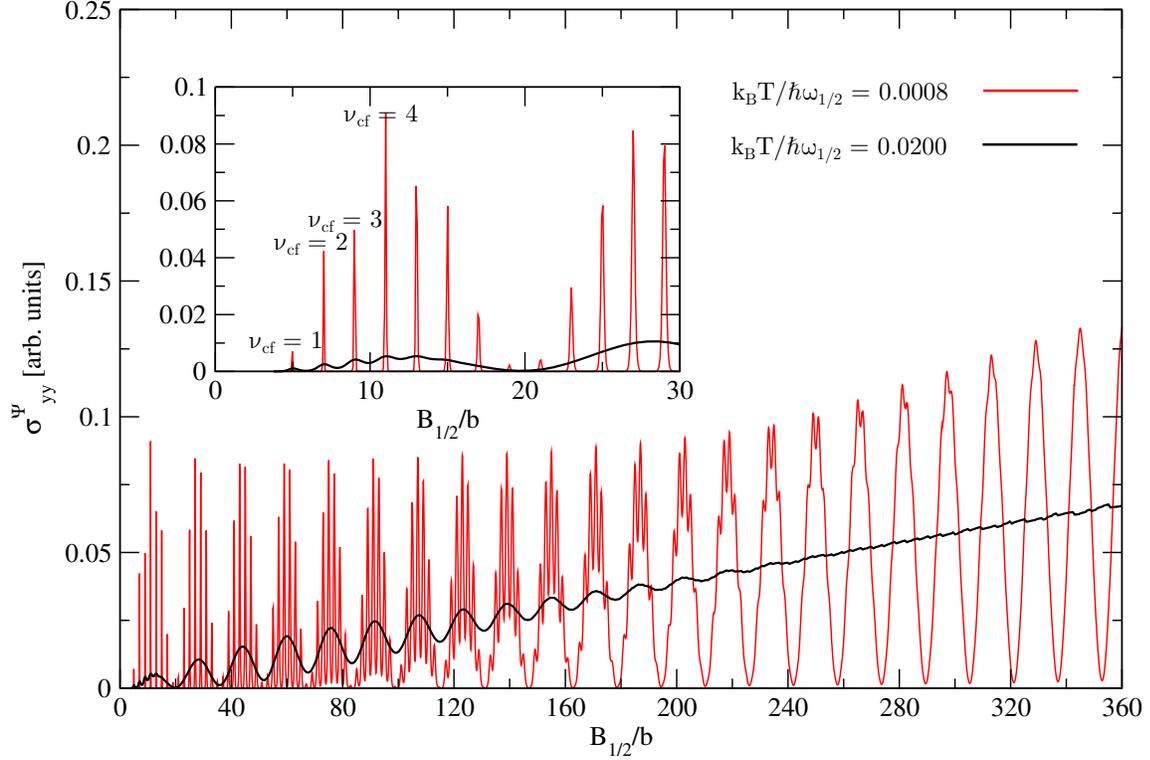}
\caption{$\sigma^\psi_{yy}$ for the Dirac composite fermions as numerically calculated using the general formula of \eqref{generalconductivity}. 
The $x$-axis is $B_{1/2}/b$ where $b$ is the effective magnetic field $b= B_{1/2} -B$ with fixed electron density $n_e = B_{1/2}/2\Phi_0$. $B_d/B_{1/2}=0.001$ where $B_d=\hbar/ed^2$ is the magnetic field associated with the modulation period $d$. Two temperatures are shown: $k_B T=0.0008\sqrt{2B_{1/2}}$ (low temperature) and $k_B T=0.02\sqrt{2B_{1/2}}$ (high temperature). At low temperatures, two types of oscillations are apparent. The longer period oscillations are the Weiss oscillations while at high effective magnetic fields (small $1/b$) the Weiss oscillations are superimposed with shorter period Shubnikov - de Haas-type (SdH) oscillations due to Landau levels crossing the chemical potential. At higher temperatures, the Weiss oscillations remain while the Shubnikov - de Haas-type oscillations are strongly suppressed. The inset shows the small $B_{1/2}/b$ region more clearly. For reference, the first four SdH-type peaks are identified with the integer filling fractions of the Dirac composite fermion Landau levels.}
\label{fig:1}
\end{figure}

In Fig. \ref{fig:1}, we numerically plot $\Delta \sigma^\psi_{yy}$ obtained from \eqref{generalconductivity} as a function of $B_{1/2}/b$ where $b$ is the effective magnetic field $b=2 \Phi_0 n_e-B$ with fixed $B_{1/2} \equiv 2 \Phi_0 n_e$. The calculation is done at two temperatures: $k_B T=0.0008\sqrt{2B_{1/2}}$ (low temperature) and $k_B T=0.02\sqrt{2B_{1/2}}$ (high temperature).

For the low temperature case (red), two types of oscillations are evident. The longer period oscillations are the Weiss oscillations while the shorter period SdH-type oscillations are superimposed on the Weiss oscillations at large magnetic fields (small $B_{1/2}/b$). The SdH-type oscillations occur when a Landau level crosses the chemical potential. Therefore, each SdH-peak can be identified with integer filling fraction $\nu_{cf}$. For reference, in the expanded inset of Fig.~\ref{fig:1}, the peaks corresponding to the first four integer filling fractions are indicated. When the temperature is increased (black), the SdH-type oscillations are strongly suppressed while the Weiss oscillations remain. This difference in robustness to temperature between the two types of oscillations is expected when $k_F d \gg 1$ \cite{peetersvasilopoulos1992scalar}. 

The contribution to the electrical resistivity can be found using \eqref{electricalconductivity}. However, to analytically exhibit the periodicity hidden in \eqref{generalconductivity}, it is helpful to approximately evaluate the expression. First, we perform the integral over $k_y$ under the assumption of weak $k_y$-dependence of the energy in the Fermi-Dirac distribution function by substituting $E_{n, k_y} = E_{n, k_y}^{(0)}$,
\begin{align}
\label{firststep}
\Delta \sigma_{yy}^{\psi} = {\tilde{V}^2 \tilde{\tau} \beta \over \hbar} \sum_{n = 0}^{\infty} {n e^{\beta E_{n, k_y}^{(0)} - \beta E_F } \over \Big(1 + e^{\beta E_{n, k_y}^{(0)} - \beta E_F}\Big)^2} e^{- z} \Big[L_{n-1}(z) - L_{n}(z) \Big]^2,
\end{align}
where we have absorbed ${\cal O}(1)$ constants into a renormalized scattering time $\tilde{\tau}$. Near half-filling where many Dirac composite fermion Landau levels are filled $n \rightarrow \infty$, it is convenient to use the asymptotic expansion of the Laguerre polynomial,
\begin{align}
\quad e^{- z/2} L_n(z) \rightarrow {\cos\Big(2 \sqrt{n z} - {\pi \over 4}\Big) \over (\pi^2 n z)^{1/4}} + {\cal O}\Big({1 \over n^{3/4}}\Big),
\end{align}
and take a continuum approximation in summing over Landau levels,
\begin{align}
\label{continuum}
n \rightarrow {E^2 \ell_b^2 \over 2 v_F^2}, \quad \sum_n \rightarrow {\ell_b^2 \over v_F^2} \int_{-\infty}^\infty E d E.
\end{align}
Substituting into \eqref{firststep},
\begin{align}
\label{integralexpression}
\Delta \sigma_{yy}^{\psi} = {\tilde{V}^2 \tilde{\tau} \beta \over \hbar}  {\ell^2_b K \over v_F^2} \int_{-\infty}^\infty d E {e^{\beta (E - E_F) } \over \Big(1 + e^{\beta (E - E_F)}\Big)^2} \sin^2\Big({E \ell^2_b K \over v_F} - {\pi \over 4}  \Big),
\end{align}
to leading order at large $n$.
The $T \rightarrow 0$ limit of the expression at finite $b$ simplifies immediately to
\begin{align}
\label{zeroTconductivity}
\Delta \sigma_{yy}^\psi (T \rightarrow 0) = {\cal A} \sin^2\Big({E_F \ell_b^2 K \over v_F} - \pi/4\Big),
\end{align}
where the non-universal coefficient ${\cal A} = {\tilde{V}^2 \tilde{\tau} \ell_b^2 K \over \hbar v_F^2}$. (Recall that $\sigma^\psi$ is the dimensionless Dirac composite fermion conductivity.) Thus, the low-temperature Dirac composite fermion conductivity exhibits oscillations as a function of $E_F \ell_b^2 K/v_F$.

If instead a periodic vector potential modulation is applied to the electronic system, the Dirac composite fermion conductivity receives a correction of the same form as \eqref{zeroTconductivity} with $\sin^2({E_F \ell^2_b K \over v_F} - \pi/4)$ replaced by $\cos^2({E_F \ell^2_b K \over v_F} - \pi/4)$.  

\section{Weiss oscillations near half-filling}

\label{weissoscillationsnearhalffilling}

\subsection{Weiss oscillations and weakly broken particle-hole symmetry}
\label{weaklybroken}

\eqref{electricalconductivity} says that the correction to the Dirac composite fermion conductivity \eqref{zeroTconductivity} produces a modulation-induced correction to the dc electrical resistivity,
\begin{align}
\label{transverse}
\Delta \rho_{xx}(T \rightarrow 0) & = {\cal A}_{xx} {e^2 \over h} \sin^2\Big({E_F \ell^2_b K \over v_F} - \pi/4\Big), \\
\label{parallel}
\Delta \rho_{yy}(T \rightarrow 0) & = - {\cal A}_{yy} {e^2 \over h} \sin^2\Big({E_F \ell^2_b K \over v_F} - \pi/4\Big),
\end{align}
where the positive constants ${\cal A}_{xx}$ and ${\cal A}_{yy}$ are proportional to ${\cal A}$ and satisfy ${\cal A}_{xx}/{\cal A}_{yy} = 1/(2 \sigma_{xx}^{\psi})^2$ at half-filling and $\tilde{V} = 0$.
In the fractional quantum Hall regime, $\rho_{xx} $ at half-filling is typically observed \cite{willett1997} to range from 100 $\Omega/\square$ to 5 $k \Omega/\square$ and so \eqref{electricalconductivity} implies $\sigma_{xx}^{\psi} < 1/16$ at $\tilde{V} = 0$. In our discussion, we concentrate on the resistivity $\Delta \rho_{xx}$ transverse to the potential modulation and will not discuss further the $\pi/2$ phase-shifted correction $|\Delta \rho_{yy}| < |\Delta \rho_{xx}|$.

The minima of $\Delta \rho_{xx}$ in \eqref{transverse} occur when
\begin{align}
\label{diracweissvector}
\ell_{b(p)}^2 = {d \over 2 k_F} \Big(p - \phi \Big), \quad p = 1, 2, 3, \ldots,
\end{align}
where we used the relation $E_F = v_F k_F$ for a linearly dispersing Dirac fermion. 
For a periodic scalar potential applied to the electronic system, (the case to which \eqref{transverse} and \eqref{parallel} apply) $\phi = -1/4$; when the electrons experience a periodic vector potential, $\phi = 1/4$. 
Naively, there is a possible minimum of \eqref{transverse} when $\ell_b^2 = d/(8k_F)$.
However, this minimum is spurious; it does not occur within the regime of validity of the continuum approximation made in \eqref{continuum}.
This can be checked by examining the exact solution plotted in Fig. \ref{fig:1}.

Evidently, both the (non-relativistic) electron gas \eqref{weissformula} and a Dirac fermion exhibit the same relation between the locations of the Weiss oscillation minima as a function of the effective magnetic field $b$, periodic potential wavelength $d$, and Fermi wave vector $k_F$. This is to be contrasted with the locations of Shubnikov - de Haas oscillation minima of a Dirac particle which are $\pi$ phase-shifted (due to the $\pi$ Berry flux enclosed by a single Dirac cone) compared to that of the electron gas \cite{Novoselov2005}. (See Appendix D of \cite{WangSenthilsecond2016} for a discussion of Dirac composite fermion Shubnikov - de Haas oscillations.)

We can translate the Dirac composite fermion parameters $\ell_b$ and $k_F$ into experimental observables. Using the relations $b = 2 \Phi_0 n_e - B$ and $k_F = \ell_B^{-1}$, \eqref{diracweissvector} becomes
\begin{align}
\label{physicalweiss}
{c \hbar \over e |2 \Phi_0 n_e - B|} = {d \over 2} \sqrt{c \hbar \over e |B|} \Big(p - \phi \Big), \quad p = 1, 2, 3, \ldots.
\end{align}
(In \cite{halperinleeread, BMF2015}, $k_F$ is instead fixed by the electron or hole density and so the factor of $\ell_B$ on the right-hand side of \eqref{physicalweiss} is replaced by $1/\sqrt{4 \pi n_e}$ or $1/\sqrt{4 \pi n_h}$.) There are two distinct ways to depart from half-filling. At fixed electron density $n_e$, we can vary $B$ and solve \eqref{physicalweiss} for the locations of the Weiss minima:
\begin{align}
\label{fixednvaryingbminima}
B^\pm(p) = 2 \Phi_0 n_e + {2 \Big(\eta \pm \sqrt{\eta^2 + 2 \eta (p - \phi)^2 d^2 \Phi_0 n_e}  \Big) \over (p - \phi)^2 d^2}
\end{align}
where $\eta = c\hbar/e$.
Pairs of minima above $B^+(p)$ and below $B^-(p)$ half-filling obey the magnetic ``sum rule,"
\begin{align}
\label{Bsumrule}
B^+(p) + B^-(p) = {4 \eta \over (p - \phi)^2 d^2} + 4 \Phi_0 n_e.
\end{align}
This magnetic ``sum rule" is consistent with the experimental findings in \cite{Kamburov2014}.
At fixed $B$, the locations of the Weiss minima as a function of the electron density occur symmetrically about half-filling:
\begin{align}
\label{fixedbvaryingnminima}
n_e^\pm(p) = {B \over 2 \Phi_0} \pm {\sqrt{\eta B} \over (p - \phi) d \Phi_0}.
\end{align}
The corresponding electric ``sum rule" becomes
\begin{align}
\label{esumrule}
n_e^+(p) + n_e^-(p) = {B \over \Phi_0}.
\end{align}
These ``sum rules" provide important experimental probes of the Dirac composite fermion theory.

\begin{figure}
\begin{center}
\includegraphics[width=1.0\columnwidth]{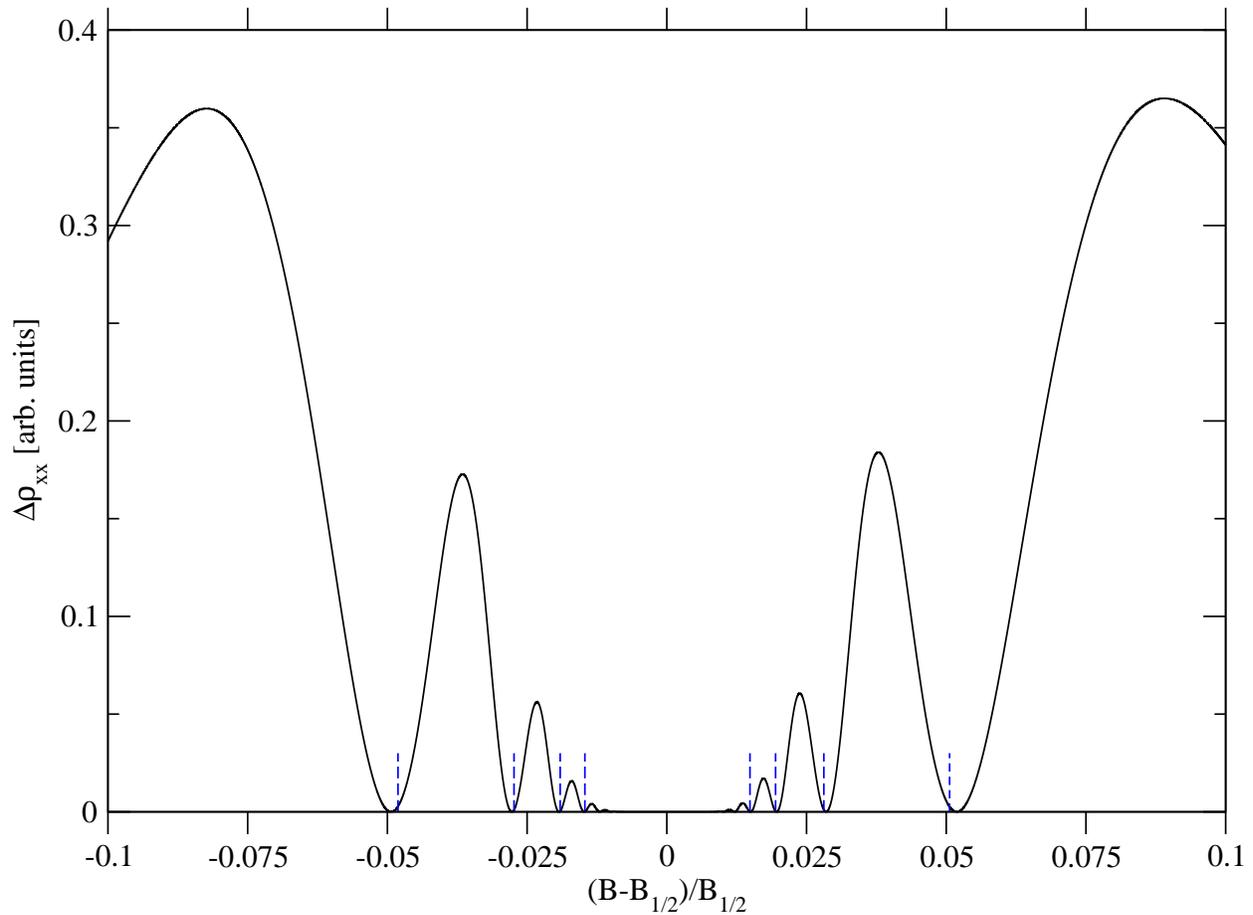}
\caption{Oscillations in $\rho_{xx}$ for electrons near half-filling as predicted by the Dirac composite fermion theory when the electron density $n_e$ is fixed and the external magnetic field is varied. Associating a magnetic field $B_d=\hbar/ed^2$ with the modulation period $d$, the result above corresponds to the choice: $B_d/B_{1/2}=0.001$, $v_F = 1$, and $k_{B}T =0.06 \sqrt{2 B_{1/2}}$. The vertical dashed bars correspond to the positions of the minima found in recent experiments \cite{Kamburov2014,LiuDengWaltz}.}
\label{fig:2}
\end{center}
\end{figure}

\begin{figure}
\begin{center}
\includegraphics[width=1.0\columnwidth]{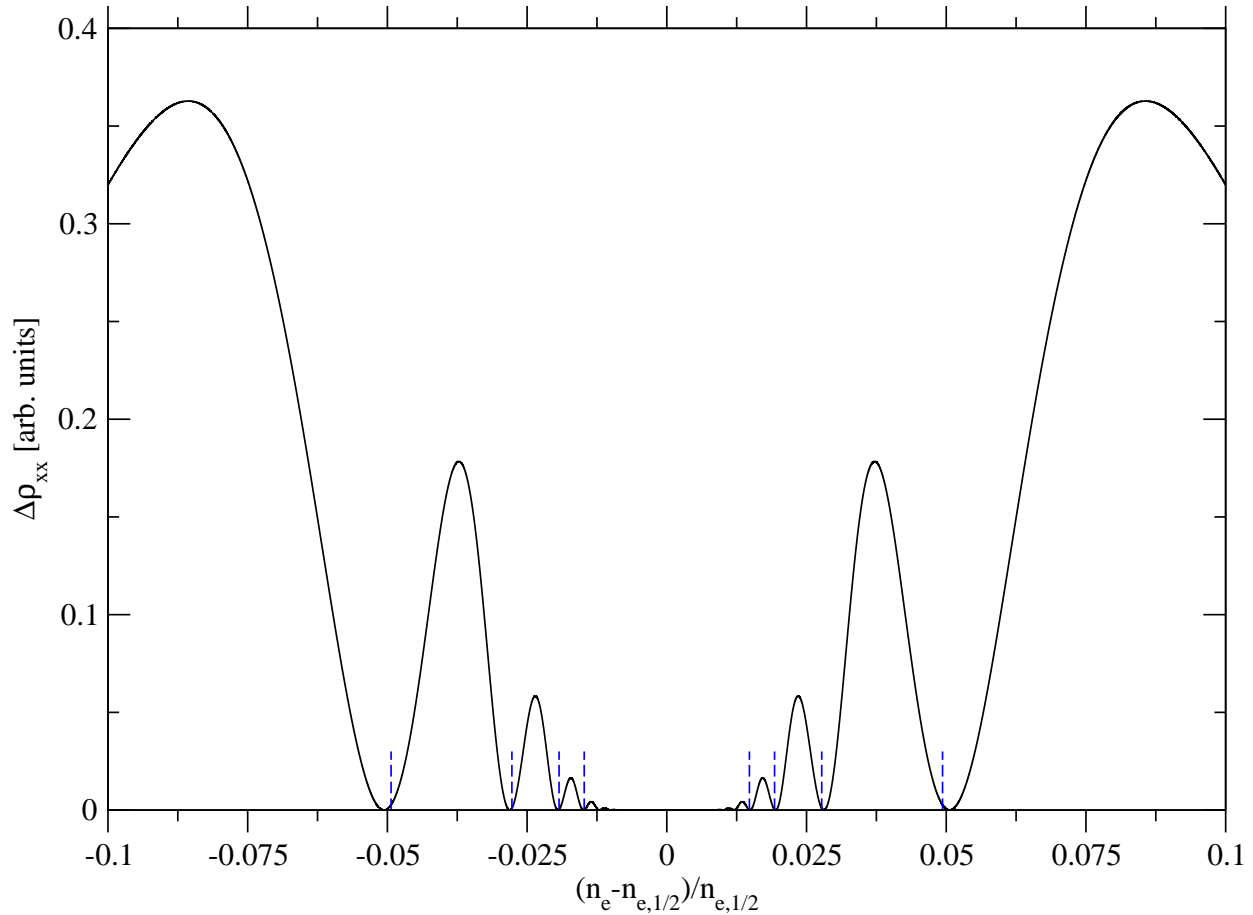}
\caption{Oscillations in $\rho_{xx}$ for electrons near half-filling as predicted by the Dirac composite fermion theory when the external magnetic field $B$ is fixed and the electron density $n_e$ is varied about its half filling value of $n_{e,1/2}=B_{1/2}/4\pi$. The result above corresponds to the choice: $B_d/B_{1/2}=0.001$, $v_F = 1$, and $k_{B}T =0.06 \sqrt{2 B_{1/2}}$. The vertical dashed bars correspond to the positions of the minima found in recent experiments \cite{Kamburov2014,LiuDengWaltz}.}
\label{fig:3}
\end{center}
\end{figure}

In Fig. \ref{fig:2}, we plot the {\it low-temperature} resistivity correction as a function of varying magnetic field and fixed electron density that is predicted by the Dirac composite fermion theory.
In Fig. \ref{fig:3}, we plot the same quantity at fixed magnetic field and varying electron density.
In both figures, the vertical dashed lines are meant to approximate to the locations of the oscillation minima found in recent experiments \cite{Kamburov2014, LiuDengWaltz}. 

In our figures, the vertical dashed lines are produced by a free, non-relativistic fermion that is electrically neutral, responds to an external scalar potential as a periodic background magnetic field, and whose density is equal to the electron density for filling fractions $\nu < 1/2$ and to the hole density for $\nu > 1/2$.
This free fermion model combines certain qualitative features of the long wavelength excitations of the composite Fermi liquid advocated in \cite{read1996cfs} with the idea that the excitations are ``electron-like" below half-filling and ``hole-like" above half-filling.

We briefly comment on how the finite-temperature curves in Figs. \ref{fig:2} and \ref{fig:3} were generated. It is necessary to generalize the $T=0$ expression for the Dirac composite fermion conductivity in \eqref{zeroTconductivity} to non-zero temperature. One effect of finite-temperature is to dampen the higher ``harmonic" minima occurring at, e.g., $p > 4$, and results in an oscillation curve that is simpler to read near $b=0$. This is straightforwardly accomplished by evaluating \eqref{integralexpression} at low temperatures ($T \ll E_F$) \cite{matulispeetersweiss2007, tahirsabeehmagneticweiss2008}:
\begin{align}
\label{finiteTgeneral}
\Delta \sigma_{yy}^\psi(T) \propto {1 \over 2} \Big[1 - {T \ell_b^2 K \over v_F \sinh({T \ell_b^2 K \over v_F})} \Big] + {T \ell_b^2 K \over v_F \sinh({T \ell_b^2 K \over v_F})}  \sin^2 \Big({E_F \ell_b^2 K \over v_F} - \pi/4 \Big).
\end{align}
Plugging \eqref{finiteTgeneral} into the relation \eqref{electricalconductivity} between the Dirac composite fermion conductivity and the electrical conductivity with a proportionality constant equal to unity (for simplicity), we obtain Figs. \ref{fig:2} and \ref{fig:3}. 

\subsection{Comparison of composite fermion theories}
\label{comparison}

In this section, we compare the Weiss oscillations expected to be produced by the Dirac composite fermion theory with the corresponding oscillations produced by the composite fermion theory of Halperin, Lee, and Read (HLR) \cite{halperinleeread} and its particle-hole conjugate, the BMF theory, introduced in \cite{BMF2015}.\footnote{We thank C. Wang, N. Cooper, B. Halperin, and A. Stern for discussions on this comparison and for pointing out an error in an earlier version of this paper.}
At short distances, these theories are distinct.
Nevertheless, to the level of resolution that current experiments probe \cite{Kamburov2014, LiuDengWaltz}, we will show that the locations of the Weiss oscillation minima produced by the three composite fermion theories agree.
This result is consistent with a recent study of the HLR theory in \cite{WangHalperin2016} and may indicate that all three theories describe the same long wavelength physics, consistent with earlier numerical work \cite{rezayi2000}.

\subsubsection{Non-relativistic composite fermion theories}

The HLR theory \cite{halperinleeread} involves a non-relativistic ``composite electron" $f$ that is minimally coupled to both a dynamical Chern-Simons gauge field $\alpha_\mu$ and the external electromagnetic gauge field $A_\mu$ with non-zero magnetic flux $\partial_x A_y - \partial_y A_x = B >0$.
The HLR Lagrangian ${\cal L}_{\rm HLR} = {\cal L}_f + {\cal L}_{\rm CS} + {\cal L}_{\rm int}$: \begin{align}
\label{bulkcompositefermionlagrangian}
{\cal L}_{f} & = f^\dagger \Big(i\partial_t + (\alpha_t + A_t) + {1 \over 2 m_f} (\partial_j - i (\alpha_j + A_j))^2 \Big) f, \cr
{\cal L}_{\rm CS} & = {1 \over 2} {1 \over 4 \pi} \epsilon^{\mu \nu \rho} \alpha_\mu \partial_\nu \alpha_\rho, \cr
{\cal L}_{\rm int} & = - {1 \over 2} \int d^2 \mathbf{r}' \Big(f^\dagger f(\mathbf{r}) - \langle f^\dagger f \rangle \Big) U_{\mathbf{r}, \mathbf{r}'} \Big(f^\dagger f(\mathbf{r}') - \langle f^\dagger f \rangle \Big),
\end{align}
where the potential $U_{\mathbf{r}, \mathbf{r}'}$ is typically taken to be either a short-ranged interaction or the long-ranged Coulomb interaction and the two spatial coordinates $\mathbf{r} = (x,y)$ and $\mathbf{r}' = (x',y')$.
The average density of composite electrons $\langle f^\dagger f \rangle$ is equal to the electron density.

The HLR theory can be viewed to result from the transformation that attaches two flux quanta to electrons near half-filling of the lowest Landau level \cite{halperinleeread}.
From this perspective and using the lowest Landau level relation between the electron density $n_e$, the hole density $n_h$, and the external magnetic field $B$ in the vicinity of half-filling, 
\begin{align}
\label{relation}
n_e + n_h = {B \over 2 \pi},
\end{align}
it is natural to consider instead the attachment of two flux quanta to the holes of the $\nu=1$ filled Landau level.
The resulting BMF theory \cite{BMF2015} involves a ``composite hole" $h$ that is minimally coupled to a dynamical gauge field $\tilde{\alpha}_\mu$ and the external electromagnetic field $A_\mu$.
The BMF Lagrangian ${\cal L}_{\rm BMF} = {\cal L}_{\rm h}  + \tilde{{\cal L}}_{\rm CS} + \tilde{{\cal L}}_{\rm int}$:
\begin{align}
\label{bulkcompositeholelagrangian}
{\cal L}_{h} & = h^\dagger \Big(i\partial_t + (\tilde{\alpha}_t - A_t) + {1 \over 2 m_h} (\partial_j - i (\tilde{\alpha}_j - A_j))^2 \Big) h, \cr
\tilde{{\cal L}}_{\rm CS} & = - {1 \over 2} {1 \over 4 \pi} \epsilon^{\mu \nu \rho} \tilde{\alpha}_\mu \partial_\nu \tilde{\alpha}_\rho + {1 \over 4 \pi} \epsilon^{\mu \nu \rho} A_\mu \partial_\nu A_\rho, \cr
\tilde{{\cal L}}_{\rm int} & = - {1 \over 2} \int d^2 \mathbf{r}' \Big(h^\dagger h(\mathbf{r}) - \langle h^\dagger h \rangle \Big) \tilde{U}_{\mathbf{r}, \mathbf{r}'} \Big(h^\dagger h(\mathbf{r}') - \langle h^\dagger h \rangle \Big),
\end{align}
where $\tilde{U}_{\mathbf{r}, \mathbf{r}'}$ parameterizes the composite hole interactions.
Notice that the electromagnetic charge of the composite holes is equal and opposite to that of the composite electrons.
This explains the opposite level of the Chern-Simons term for $\tilde{\alpha}_\mu$ in $\tilde{\cal L}_{\rm CS}$.
The filled Landau level vacuum is represented by the second term in $\tilde{{\cal L}}_{\rm CS}$ which ensures an integer contribution to the electrical Hall conductivity.
The average density of composite holes $\langle h^\dagger h \rangle$ is equal to the hole density, $n_h = {B \over 2 \pi} - n_e$.

We simplify the analysis of the HLR and BMF theories by ignoring fluctuations of the emergent gauge fields.
This approximation is analogous to the approximation that was made in our study of the Dirac composite fermion theory.
This approximation enables us to ignore the composite electron and composite hole interaction terms in ${\cal L}_{\rm int}$ and $\tilde{\cal L}_{\rm int}$ upon use of the equation of motion of the temporal component of the respective emergent gauge field.

In contrast to the Dirac composite fermion, the composite electrons and composite holes directly couple to the electromagnetic gauge field.
These couplings have an important consequence: within the HLR and BMF theories, a periodic electromagnetic scalar potential induces both a periodic scalar and periodic vector potential background of fixed relative magnitude for the composite fermions.
To see how this occurs in the HLR theory, we notice that $\alpha_t$ functions as a Lagrange multiplier that imposes the constraint:
\begin{align}
\label{constraint}
f^\dagger f = - {1 \over 4 \pi} \beta,
\end{align}
where $\beta = \partial_x \alpha_y - \partial_y \alpha_x$ is the magnetic flux of the emergent gauge field.
Coupled with the relation $\langle f^\dagger f \rangle = n_e$, this constraint leads to the usual conclusion that the emergent magnetic flux $\beta$ completely screens, on average, the external magnetic flux at half-filling.
When a periodic electromagnetic scalar potential $A_t(\mathbf{r})$  of the form \eqref{scalarpotential} is applied to the system, a modulation in the composite electron density is produced, 
\begin{align}
\delta \langle f^\dagger f \rangle = {m_f \over 2 \pi} A_t(\mathbf{r}).
\end{align}
The coefficient $m_f/2\pi$ is the compressibility of a free (composite) fermion gas in two dimensions.
\eqref{constraint} implies a corresponding modulation of the emergent magnetic field $\delta \beta = -2 m_f A_t(\mathbf{r})$.
Identical considerations apply to $\tilde{\alpha}_t$ in the BMF theory and result in a periodic magnetic flux $\tilde{\beta} = - 2 m_h A_t(\mathbf{r})$ where $\tilde{\beta} =  \partial_x \tilde{\alpha}_y - \partial_y \tilde{\alpha}_x$.

\subsubsection{Weiss oscillations of non-relativistic composite fermion theories}

We can borrow the analysis in \cite{BMF2015} that determined the expected correction to the electrical resistivity when the composite electrons and composite holes are subject to both a periodic scalar and vector potential.
More careful treatments of the HLR theory, which can be readily adapted to BMF theory, can be found in \cite{mirlin1998cfs, vonoppensternhalperin1998, zwerschkegerhardts1999}.
For the composite electrons, 
\begin{align}
\label{compositeelectronresistivity}
\delta \rho_{xx} \Big|_f & \propto \Big[{B_{\rm eff} \over |B_{\rm eff}|} \cos\Big(k^{\rm f}_F \ell_{B_{\rm eff}}^2 K - {\pi \over 4}\Big) + {2 k^{\rm f}_F \over K} \sin\Big(k^{\rm f}_F \ell_{B_{\rm eff}}^2 K - {\pi \over 4}\Big) \Big]^2 \cr
& \propto \Big[\sin\Big(k_F^f \ell^2_{\rm B_{\rm eff}} K - {\pi \over 4} + {B_{\rm eff} \over |B_{\rm eff}|}\tan^{-1}({K \over 2 k_F^f})  \Big) \Big]^2 ,
\end{align}
where $B_{\rm eff} = B - 2 \Phi_0 n_e$, $k^{\rm f}_F = \sqrt{4 \pi n_e}$, $\ell^{-2}_{\rm B_{\rm eff}} = e|B_{\rm eff}|/c\hbar$, and the periodic modulation wave vector $K = 2\pi/d$.
The first term in \eqref{compositeelectronresistivity} is a result of the periodic scalar potential, while the second term arises from the periodic vector potential.
The notation $\Big|_f$ reminds us that $\delta \rho_{xx} \Big|_f$ is the contribution to the electrical resistivity arising from the HLR theory of composite electrons.
For composite holes, the leading correction to the resistivity,
\begin{align}
\label{compositeholeresistivity}
\delta \rho_{xx} \Big|_h & \propto \Big[- {B_{\rm eff} \over |B_{\rm eff}|} \cos \Big(k^{\rm h}_F \ell_{B_{\rm eff}}^2 K - {\pi \over 4}\Big) + {2 k^{\rm h}_F \over K} \sin\Big(k^{\rm h}_F \ell_{B_{\rm eff}}^2 K - {\pi \over 4}\Big)  \Big]^2 \cr
& \propto \Big[\sin\Big(k_F^h \ell^2_{\rm B_{\rm eff}} K - {\pi \over 4} - {B_{\rm eff} \over |B_{\rm eff}|}\tan^{-1}({K \over 2 k_F^h})  \Big) \Big]^2 ,
\end{align}
where $k_F^{\rm h} = \sqrt{4\pi n_h} = \sqrt{4 \pi ({B \over 2 \pi} - n_e)}$.
The opposite electromagnetic charge of the composite holes, relative to the composite electrons, results in the $\pi$-phase shift between the contributions of the scalar and vector potential modulations to the resistivity.
Because $k_F^{{\rm f, h}}/K \gg 1$ in experiment, the scalar contribution to the resistivity correction was ignored in \cite{BMF2015}.
In the next section, we will show that if this correction is included, the locations for the Weiss oscillation minima produced by the Dirac composite fermion, HLR, and BMF theories coincide in the experimental regime.

\subsubsection{Oscillation minima comparison}


First consider fixed electron density $n_e > 0$ and varying the magnetic field about half-filling $B = 2 \Phi_0 n_e$. 
Parameterizing the deviation away from the minima of the Dirac composite fermion theory  $B^\pm(p)$ given in \eqref{fixednvaryingbminima} by the dimensionless function $\epsilon^\pm(p)/(d \sqrt{n_e})^3$, we substitute
\begin{align}
B = B^\pm(p) + 2 \Phi_0 n_e {\epsilon^\pm(p) \over d^3 n_e^{3/2}}
\end{align}
into \eqref{compositeelectronresistivity} and \eqref{compositeholeresistivity} for the HLR and BMF theories to find:
\begin{align}
\delta \rho_{xx} \Big|_f \propto \delta \rho_{xx} \Big|_h \propto  \Big[\sin\Big(\pi p \mp {(1 + 4 p)^2 \pi^{3/2} \over 16} {\epsilon^\pm(p) \over d^2 n_e} + {1 \over 2 + 8 p} {1 \over d^2 n_e} \Big) \Big]^2,
\end{align}
where we have only retained the leading correction in the $d^2 n_e \rightarrow \infty$ limit. 
Remarkably, the oscillation minima for the HLR and BMF theories coincide and deviate from the Dirac composite fermion result by ${2 \Phi_0 n_e \over d^3 n_e^{3/2}} \epsilon^\pm(p)$ where
\begin{align}
\label{deviation}
\epsilon^\pm(p) = \pm \Big({2 \over (1 + 4 p) \pi^{1/2}} \Big)^3.
\end{align}
The HLR and BMF minima are shifted outward from the value of the magnetic field at half-filling, $2 \Phi_0 n_e$.
In particular, for a given Dirac composite fermion oscillation minimum labeled by $p$, the HLR and BMF minima are shifted away from the dashed lines in Fig. \ref{fig:2}.

Next consider fixed magnetic field and varying the electron density about half-filling.
We substitute 
\begin{align}
n_e = n_e^{\pm}(p) + {B \over 2 \Phi_0} {\epsilon^\pm(p) \over d^3 (B/2 \Phi_0)^{3/2}}
\end{align}
into \eqref{compositeelectronresistivity} and \eqref{compositeholeresistivity} for the HLR and BMF theories to find:
\begin{align}
\delta \rho_{xx} \Big|_f \propto \delta \rho_{xx} \Big|_h \propto  \Big[\sin\Big(\pi p \mp {(1 + 4 p)^2 \pi^{3/2} \over 16} {\epsilon^{\pm}(p) \over d^2 (B/2 \Phi_0)} + {1 \over (2 + 8 p)} {1 \over d^2 B} \Big) \Big]^2,
\end{align}
where we have only retained the leading correction in the $d^2 B/\Phi_0 \rightarrow \infty$ limit.
As in the previous case, the oscillation minima for the HLR and BMF theories coincide and the deviation from the Dirac composite fermion result is parameterized by $\epsilon^\pm(p)$ given in \eqref{deviation}.
The oscillation minima occur symmetrically about the value of the electron density at half-filling and are shifted ``outward," away from $B/2\Phi_0$, relative to the values of the Dirac composite fermion theory.

We characterize the deviation between the oscillation minima produced by the Dirac composite fermion and those expected of the HLR or BMF theories using the ratio,
\begin{align}
\label{ratio}
\Big| {2 \Phi_0 n_e {\epsilon^\pm(p) \over d^3 n_e^{3/2}} \over B^{\pm}(p)} \Big| \approx \Big| {\epsilon^{\pm}(p) \over d^3 n_e^{3/2}} \Big|.
\end{align}
The difference between the Dirac composite fermion theory and HLR or BMF theory is magnified as $d^2 n_e$ is reduced.
At fixed electron density, as the external modulation period $d$ is reduced, the Weiss oscillation minima begin to overlap with the quantum oscillation minima reflective of nearby quantum Hall states and may be challenging to identify experimentally.
(Recall that the locations of the Weiss oscillation minima as a function of varying external magnetic field are generally proportional to a power of $k_F/d$.)

We estimate the ratio in \eqref{ratio} using $n_e \sim 1.75 \times 10^{11}\ {\rm cm}^{-2}$ and $d \sim 200\ {\rm nm}$, consistent with the experiments in \cite{Kamburov2014, LiuDengWaltz}:
\begin{align}
\Big| {\epsilon^{\pm}(p) \over d^3 n_e^{3/2}} \Big| \sim {2.5 \times 10^{-3} \over (1 + 4 p)^3}.
\end{align}
Evidently, in the experimental regime, the various composite fermion theories agree within $.002 \%$ accuracy, but, unfortunately, do not yet explain the results in \cite{Kamburov2014, LiuDengWaltz}.

\section{Discussion}
\label{discussion}

In this article, we studied the consequences of weakly-broken particle-hole symmetry for the Weiss oscillations about half-filling using the Dirac composite fermion theory introduced by Son \cite{Son2015}. Within this description, the fact that the density of excitations contributing to the quantum oscillations is fixed by the applied magnetic field has important implications. At fixed electron density, the locations of Weiss oscillation minima as a function of applied magnetic field obey a magnetic ``sum rule" that is {\it not} symmetric about half-filling. This ``sum rule" is qualitatively consistent with the observations made in \cite{Kamburov2014}. At fixed magnetic field, the Weiss minima occur symmetrically about the half-filling as the electron density is varied.
Further experimental investigations comparing magnetic and electric ``sum rules" could shed additional light on the dynamics at half-filling. 

We also compared the Weiss oscillations expected to be produced by the Dirac composite fermion theory with the corresponding oscillations produced by the composite fermion theory of Halperin, Lee, and Read \cite{halperinleeread} and its particle-hole conjugate \cite{BMF2015}.
At short distances, these theories are distinct.
Nevertheless, in the regime of parameter space probed by current experiments \cite{Kamburov2014, LiuDengWaltz}, we showed that the locations of the Weiss oscillation minima produced by the various composite fermion theories coincide.
This result is consistent with a recent study of the HLR theory in \cite{WangHalperin2016} and may indicate that all three theories describe the same long wavelength physics, consistent with earlier numerical work \cite{rezayi2000}.

It would be interesting to further examine the quantum oscillations about other even-denominator filling fractions of the two-dimensional electron gas. In \cite{Shahar1996}, a careful study of the symmetries of resistance curves near the transition between filling fractions $\nu=1/3 \rightarrow 0$ revealed an emergent composite fermion particle-hole symmetry. It is interesting to speculate whether this emergent symmetry is supported by quantum oscillations.

The Dirac composite fermion theory has an obvious generalization to the half-filled zeroth Landau level of graphene. It would be interesting to investigate both theoretically and experimentally the nature of Weiss oscillations in these systems. For example, if present, how do the oscillations depend on the dominant interactions in these systems?

In \cite{BalramRifmmodeCsabaJain2015}, it was found using a model wave function for composite fermions \cite{jainCF} that the Fermi wave vector of the excitations near half-filling, as inferred from Friedel oscillations, appears ``electron-like" for $\nu < 1/2$ and ``hole-like" for $\nu > 1/2$, in qualitative agreement with the experiments in \cite{Kamburov2014, LiuDengWaltz}. In contrast, we have discussed the implications of a particle-hole symmetric {\it field theoretic} description of the state at half-filling which asserts that the Fermi wave vector is equal to the inverse magnetic length $\ell^{-1}_B$. It would be beneficial to have a better understanding of the relation between these two approaches. 

The primary simplification used in determining the Weiss oscillations of the Dirac composite fermion theory was to ignore the fluctuations of the emergent gauge field. In \cite{KMTW2015}, it was observed that Coulomb interactions between electrons softened the strong correlations inherent in the Dirac composite fermion theory. This result can be interpreted to reinforce the expectation that Coulomb interactions are an essential ingredient to the physics at half-filling. Thus, Weiss oscillations provide an invaluable probe of the strong correlations inherent in these systems.

In recent work, Levin and Son \cite{LevinSon2016} discuss an exact relation between the Hall conductivity and susceptibility that appears to distinguish composite fermion theories.
Further comparison of the various composite fermion theories constitutes important work for the future.

\section*{Acknowledgments}
We thank D. Goldhaber-Gordon for useful correspondence and M. Shayegan for discussions, for helpful comments on an early draft of the paper, and for sharing with us some of his group's on-going work. 
We thank C. Wang, N. Cooper, A. Stern, and especially B. Halperin for pointing out an error in an earlier version of this paper and for discussions about the Weiss oscillations of various composite fermion theories..
M.M. thanks the participants of the UCR condensed matter lunch club for their comments and questions. This research was supported in part by the University of California (M.M.), the DOE Office of Basic Energy Sciences, contract DE-AC02-76SF00515 (S.R. and A.C.), and a NSERC PGS-D Scholarship (A.C.). M.M. is grateful for the generous hospitality of the Kavli Institute for Theoretical Physics NSF PHY-11-25915.

\bibliography{diraccfqoscillations}{}
\bibliographystyle{utphys}

\end{document}